\documentclass[sigconf]{acmart}
\usepackage{pifont} % checkmark

\AtBeginDocument{%
  \providecommand\BibTeX{{%
    \normalfont B\kern-0.5em{\scshape i\kern-0.25em b}\kern-0.8em\TeX}}}

\copyrightyear{2024}
\acmYear{2024}
\setcopyright{rightsretained}
\acmConference[CHI '24]{Proceedings of the CHI Conference on Human Factors in Computing Systems}{May 11--16, 2024}{Honolulu, HI, USA} \acmBooktitle{Proceedings of the CHI Conference on Human Factors in Computing Systems (CHI '24), May 11--16, 2024, Honolulu, HI, USA} \acmDOI{10.1145/3613904.3641896}
\acmISBN{979-8-4007-0330-0/24/05}

\begin{document}

\title{Sketching AI Concepts with Capabilities and Examples: \\ AI Innovation in the Intensive Care Unit}

\settopmatter{authorsperrow=4}

\author{Nur Yildirim}
\affiliation{%
  \institution{Carnegie Mellon University}
  \city{Pittsburgh}
  \state{PA}
  \country{USA}}
\email{yildirim@cmu.edu}

\author{Susanna Zlotnikov}
\affiliation{%
  \institution{Carnegie Mellon University}
  \city{Pittsburgh}
  \state{PA}
  \country{USA}}

\author{Deniz Sayar}
\affiliation{%
  \institution{Izmir University of Economics}
  \city{Izmir}
  % \state{PA}
  \country{Turkey}}

\author{Jeremy M. Kahn}
\affiliation{%
  \institution{University of Pittsburgh}
  \city{Pittsburgh}
  \state{PA}
  \country{USA}}
% \email{jeremykahn@pitt.edu}

\author{Leigh A. Bukowski}
\affiliation{%
  \institution{University of Pittsburgh}
  \city{Pittsburgh}
  \state{PA}
  \country{USA}}

\author{Sher Shah Amin}
\affiliation{%
  \institution{University of Pittsburgh}
  \city{Pittsburgh}
  \state{PA}
  \country{USA}}

\author{Kathryn A. Riman}
\affiliation{%
  \institution{University of Pittsburgh}
  \city{Pittsburgh}
  \state{PA}
  \country{USA}}

\author{Billie S. Davis}
\affiliation{%
  \institution{University of Pittsburgh}
  \city{Pittsburgh}
  \state{PA}
  \country{USA}}

\author{John S. Minturn}
\affiliation{%
  \institution{University of Pittsburgh}
  \city{Pittsburgh}
  \state{PA}
  \country{USA}}

\author{Andrew J. King}
\affiliation{%
  \institution{University of Pittsburgh}
  \city{Pittsburgh}
  \state{PA}
  \country{USA}}

\author{Dan Ricketts}
\affiliation{%
  \institution{University of Pittsburgh}
  \city{Pittsburgh}
  \state{PA}
  \country{USA}}

\author{Lu Tang}
\affiliation{%
  \institution{University of Pittsburgh}
  \city{Pittsburgh}
  \state{PA}
  \country{USA}}

\author{Venkatesh Sivaraman}
\affiliation{%
  \institution{Carnegie Mellon University}
  \city{Pittsburgh}
  \state{PA}
  \country{USA}}

\author{Adam Perer}
\affiliation{%
  \institution{Carnegie Mellon University}
  \city{Pittsburgh}
  \state{PA}
  \country{USA}}

\author{Sarah M. Preum}
\affiliation{%
  \institution{Dartmouth College}
  \city{Hanover}
  \state{NH}
  \country{USA}}

\author{James McCann}
\affiliation{%
  \institution{Carnegie Mellon University}
  \city{Pittsburgh}
  \state{PA}
  \country{USA}}
% \email{jmccann@cs.cmu.edu}

\author{John Zimmerman}
\affiliation{%
  \institution{Carnegie Mellon University}
  \city{Pittsburgh}
  \state{PA}
  \country{USA}}
% \email{johnz@cs.cmu.edu}

\renewcommand{\shortauthors}{Yildirim, et al.}

\begin{abstract}
Advances in artificial intelligence (AI) have enabled unprecedented capabilities, yet innovation teams struggle when envisioning AI concepts. Data science teams think of innovations users do not want, while domain experts think of innovations that cannot be built. A lack of effective ideation seems to be a breakdown point. How might multidisciplinary teams identify buildable and desirable use cases? This paper presents a first hand account of ideating AI concepts to improve critical care medicine. As a team of data scientists, clinicians, and HCI researchers, we conducted a series of design workshops to explore more effective approaches to AI concept ideation and problem formulation. We detail our process, the challenges we encountered, and practices and artifacts that proved effective. 
\textcolor{black}{We discuss the research implications for improved collaboration and stakeholder engagement, and discuss the role HCI might play in reducing the high failure rate experienced in AI innovation.}

\end{abstract}

\begin{CCSXML}
<ccs2012>
<concept>
<concept_id>10003120.10003123.10010860</concept_id>
<concept_desc>Human-centered computing~Interaction design process and methods</concept_desc>
<concept_significance>500</concept_significance>
</concept>
</ccs2012>
\end{CCSXML}

\ccsdesc[500]{Human-centered computing~Interaction design process and methods}

\keywords{Brainstorming, ideation, human-centered AI, healthcare}


\maketitle

\section{Introduction}
\textcolor{black}{Artificial intelligence (AI) is transforming the landscape of healthcare. From cancer diagnosis \cite{cai2019hello} to prognosis \cite{yang2019unremarkable}, automated documentation \cite{king2023voice}, and treatment recommendations \cite{sivaraman2023ignore}, AI applications in healthcare offer the promise of improved clinician experience and better healthcare outcomes for patients. While AI’s technical advances showcase impressive performance in lab settings, AI systems largely fail when moving to clinical practice \cite{yang2016investigating, osman2021realizing, yu2018artificial, galsgaard2022artificial, thieme2023designing}. HCI researchers note that the clinical utility and actionability –whether clinicians can take specific actions based on a prediction– of healthcare AI applications often remain unclear \cite{ghosh2023framing, yildirim2021technical, thieme2020machine}. Clinicians do not use AI systems, often because systems do not deliver what they need.}

\textcolor{black}{The challenge of making AI advances useful in real world contexts is not unique to healthcare. Today, the majority of AI initiatives fail, as they fail to generate enough value for users or for service providers \cite{weiner2020ai, joshi2021so, ermakova2021beyond}. Product teams share experiencing repeated AI failures due to selecting and working on the wrong problem – high-risk projects that may or may not be valuable or that entail unavoidable challenges around fairness and bias \cite{yildirim2023investigating, holstein2019improving, passi2019problem, boyarskaya2020overcoming}. AI development practices remain technology-driven with little attention to human needs and wants \cite{yildirim2023investigating}. Stakeholders that do not have a background in data science or AI are rarely involved in conversations around the objective of the underlying model or the overall problem formulation, if involved at all \cite{feffer2023preference, delgado2023participatory}. Challenges in multidisciplinary collaboration across team members poses a major barrier to AI design and development \cite{piorkowski2021ai, kross2021orienting, passi2019problem, nahar2022collaboration}. As AI capabilities become readily available, a critical question arises: How can multidisciplinary innovation teams effectively identify low-risk, high-value AI use cases?}

\textcolor{black}{In response to these challenges, HCI researchers called for human-centered, participatory approaches to AI development –especially in early ideation and problem formulation phases– to reduce the risk of developing unwanted technology \cite{delgado2023participatory, yildirim2023investigating, yang2020re}. Studies on industry best practices revealed that effective innovation teams brainstorm using AI capabilities and examples of AI applications to close knowledge gaps between data science, HCI, and domain expertise \cite{yang2018investigating, yildirim2022experienced, yildirim2023investigating}. These emergent AI innovation practices resembled a blend of user-centric and tech-centric approaches, where teams rapidly generated many AI concepts to match \cite{bly1999design} AI capabilities with human needs within a specific product domain \cite{yildirim2023investigating}. An emerging body of research have started to explore how team members and domain stakeholders might envision and co-design AI use cases (e.g., in law \cite{delgado2022uncommon}, public services \cite{stapleton2022imagining}, accessibility \cite{valencia2023less}), and the types of design process processes, tools, and methods that might prove effective \cite{murray2022metaphors, kuo2023understanding, dove2020monsters}.} 

\textcolor{black}{Building on this line of research, we set out to explore clinically relevant and feasible AI uses cases for intensive care within a multidisciplinary team of AI researchers, HCI researchers, data scientists, and healthcare professionals. Our prior work detailed the development of \textit{the AI Brainstorming Kit} \cite{yildirim2023creating} – a resource to help HCI experts facilitate AI concept ideation within multidisciplinary teams, especially to identify low-risk, high-value concepts where \textit{moderate AI performance} can create value. In this paper, we present a reflective account of our design process as a case study of early phase AI innovation, with a specific focus on capturing the iterative research activities. Our team had access to a rich ICU dataset (similar to MIMIC \cite{johnson2019mimic}) that was collected across 39 intensive care units (ICUs) from 18 hospitals. We engaged in an iterative design process to broadly explore the problem-opportunity space for \textit{getting the right design} \cite{buxton2010sketching}. We conducted a three-phase study, where we moved from ideation to problem formulation, concept design, and initial assessment with end users.}

\textbf{Phase 1 Brainstorming} focused on envisioning many AI concepts and use cases for the ICU, before selecting and building an application. We conducted two brainstorming workshops within our multidisciplinary team. The first workshop followed a traditional user-centered design approach with a focus on user needs. The second workshop combined user-centered and matchmaking \cite{bly1999design} approaches to consider both user needs \textit{and} AI capabilities simultaneously. Building on \textcolor{black}{\textit{the AI Brainstorming Kit \cite{yildirim2023creating}}}, we used a set of AI capabilities and examples to scaffold ideation, selecting examples where moderate model performance was `good enough’ to produce value. An assessment of outputs from each workshop demonstrated that the latter approach resulted in more effective brainstorming with many concepts that were low-risk in terms of feasibility and clinician acceptance, and medium to high-value for clinicians.

\textbf{Phase 2 Problem Formulation} focused on detailing a subset of AI concepts further \textit{(e.g., predicting medication availability and anticipatory ordering)}. Our brainstorming sessions yielded many concepts that leveraged AI capabilities in ways that provided utility for clinicians; however, it was unclear whether and how these could operationalize our unique ICU dataset. To tackle this challenge, we conducted a follow-up workshop session, where we detailed the required model performance, point of interaction, data requirements, and risks (e.g., consequences of potential errors) for 12 use cases we previously identified. We created a worksheet detailing the data, model reasoning, and interaction form to disentangle interaction design and model building considerations. This proved an effective artifact for refining concepts, revealing unreliable data, and considering if simpler versions of a concept might also be valuable.

\textbf{Phase 3 Sketching and Co-Design} explored further refining an AI concept towards prototyping to elicit early phase user feedback. We selected a concept that aimed to predict if a patient is eligible to receive the protocol for assessing readiness for liberation from mechanical ventilation. We created sketches detailing the concept. We conducted four co-design workshops with 11 clinicians to probe whether and how this concept might support them in considering and executing this specific evidence-based protocol. Participants perceived the concept as valuable, and they articulated detailed design requirements for interaction design as well as model building and data.

\textcolor{black}{This paper makes two contributions. First, we present a rare case study of early phase AI innovation within a multidisciplinary team of data scientists, domain experts, and HCI researchers. We describe the challenges faced across brainstorming, concept design, and initial assessment. These practices serve as a starting point for multidisciplinary teams to structure design activities for navigating early phase human-centered AI innovation. Second, we discuss remaining challenges and outline opportunities for HCI researchers to better support and facilitate effective collaboration and stakeholder engagement in AI innovation projects, specifically to identify low-risk, high-value use cases in high-stakes contexts including healthcare and beyond.}

\vspace{-4mm}

\section{Related Work}
% We first provide a brief description of types of ideation methods and processes in HCI. Next, we situate our work in the broader context of human-centered AI, collective ideation and problem formulation. Finally, we briefly touch on AI innovation in healthcare and ICUs.

\subsection{\textcolor{black}{Challenges of AI Product Innovation}}

\textcolor{black}{HCI research characterizes \textit{technology as a material} \cite{wiberg2014methodology, redstrom2005technology} that designers can explore to envision novel interactions (e.g., bluetooth \cite{sundstrom2011inspirational}, haptics \cite{moussette2010designing}, and software \cite{ozenc2010support}). Embracing this material lens, researchers have framed `AI as design material’ to explore AI’s opportunities and challenges for HCI research and practice \cite{yang2020re, subramonyam2022solving, yildirim2022experienced}. From language translation to text summarization, medical diagnosis and image generation, technical AI advances offer unprecedented capabilities. While these advances open up a novel space for interactive systems, they also pose unique challenges to designing AI products and services \cite{yang2020re}. A large body of research investigated the challenges around explainability \cite{liao2020questioning, abdul2018trends}, trust and reliance \cite{buccinca2021trust}, user control \cite{shneiderman2020human}, feedback \cite{stumpf2007toward}, error recovery \cite{kocielnik2019will}, and fairness-related harms \cite{veale2018fairness}, just to name a few.}

\textcolor{black}{While research efforts have largely focused on mitigating issues that arise post-deployment, recent research points to a more consequential problem: more than 85\% of AI innovation projects fail pre-deployment \cite{weiner2020ai, joshi2021so, ermakova2021beyond}. Failure includes taking on projects that are too complex or infeasible; selecting problems that entail unavoidable fairness issues, such as privacy concerns or algorithmic bias; and building systems that in the end fail to generate enough value for customers or service providers \cite{weiner2020ai}. Some researchers critiqued this breakdown from a perspective of `validity’, raising the importance of asking whether an AI system provides any benefits in the first place \cite{raji2022fallacy}. Studies investigating industry practices attribute AI failures to lack of human-centered approaches and ineffective collaboration between cross-disciplinary team members in early problem formulation phases of a project \cite{piorkowski2021ai, mao2019data, kross2021orienting, nahar2022collaboration, yildirim2023investigating, deng2023investigating, varanasi2023currently, wang2023designing}.}

\textcolor{black}{In recent years, resources in the form of guidelines and toolkits became available to address some of AI’s design challenges (e.g., human-AI guidelines \cite{amershi2019guidelines, apple2019, google2019pair}, fairness toolkits \cite{deng2022exploring}). However, investigations on how teams use these resources indicate that these resources mainly help at later stages, \textit{after} problem selection and formulation. Practitioners ask for resources that support early phase ideation and problem formulation to discover use cases where AI might be a good solution \cite{yildirim2023investigating}. A related strand of research investigating industry best practices revealed that effective innovation teams work with AI capabilities and examples to scaffold cross-disciplinary ideation \cite{yildirim2022experienced, yang2018investigating, yildirim2023investigating, zimmerman2020ux}. These resources detail what AI can do instead of how AI works using non-technical terms (e.g., detect customer patterns; predict seasonality trends), which seem to help user experience designers and product managers gain a practical understanding of AI \cite{yildirim2023investigating}.}

\textcolor{black}{Finally, researchers report limitations of user-centered design (UCD) in AI innovation \cite{forlizzi2018moving, zdanowska2022study, girardin2017user, yildirim2023investigating, yang2018mapping, oppermann2020data}, and highlight emergent design processes that blend UCD and \textit{matchmaking} \cite{bly1999design} – an innovation process that starts with a technical capability to systematically search for customers that might benefit from it. Researchers also point out that innovation teams often focus on complex use cases where near-perfect AI performance is needed for a concept to be useful \cite{feng2023ux, yang2020re}. A recent analysis of 40 AI applications note that the majority of real-world applications in fact leverage moderate model performance, suggesting that teams should focus on cases where imperfect AI can create value \cite{yildirim2023creating}.}

\textcolor{black}{HCI researchers have explored AI concept ideation through design-led inquiry to provide first-person accounts of their design process, challenges, and emerging solutions \cite{yang2019sketching, kayacik2019identifying, benjamin2021machine, zimmerman2022recentering, khanshan2023case}. For example, Yang et al. detailed how a team of HCI and NLP researchers envisioned and prototyped AI-powered features for Microsoft Word \cite{yang2019sketching}. Kayacik et al. described how UX designers and AI research scientists envisioned AI-driven concepts using generative AI capabilities for music creation \cite{kayacik2019identifying}. In the same spirit, we set out to contribute a detailed case study of our ideation process for envisioning and designing AI use cases for intensive care.}

\subsection{\textcolor{black}{Broadening Participation in AI Design}}

\textcolor{black}{A growing body of work has called for socio-technical, participatory approaches to meaningfully engage domain stakeholders throughout the AI development lifecycle \cite{delgado2021stakeholder, delgado2023participatory, zhang2023deliberating, bell2023think, tahaei2023human, cooper2022systematic}. Prior research notes that stakeholders with little to no background in data science or AI are rarely involved in problem selection and formulation, if involved at all \cite{kawakami2022improving, holten2020shifting, delgado2023participatory, feffer2023preference}. There is a knowledge gap between data science and domain expertise \cite{yang2019sketching, kross2021orienting, yildirim2023investigating}: Domain experts and designers struggle to understand what AI can do, they often envision AI services that cannot be built \cite{dove2017ux, yang2020re, yu2020keeping, liao2023designerly}. Data scientists find it challenging to elicit needs from domain experts, and without this input, they tend to envision AI services that users and impacted stakeholders do not want \cite{piorkowski2021ai, mao2019data, kross2021orienting, nahar2022collaboration}. Teams do not seem to ideate; they focus on building a single application without exploring the space of possibilities \cite{yildirim2023investigating}.}

\textcolor{black}{Recent HCI research has proposed new design methods, artifacts, and resources, such as metaphors \cite{dove2020monsters, murray2022metaphors}, AI lifecycle comicboarding \cite{kuo2023understanding}, onboarding materials \cite{cai2021onboarding}, and other artifacts \cite{ayobi2023computational, lam2023model} to facilitate effective stakeholder engagement. Notably, research employing this type of resources often focuses on later stage AI phases, detailing how to refine existing AI systems or mitigate harmful outputs. Relatively little research has offered a detailed account of early phase ideation and problem formulation with domain experts and impacted stakeholders. Few examples worth noting present case studies on envisioning and designing AI use cases in child welfare \cite{stapleton2022imagining}, fact-checking \cite{liu2023human}, law \cite{delgado2022uncommon}, and content moderation \cite{halfaker2020ores}, and accessibility \cite{valencia2023less, morrison2017imagining}. We draw on this strand of research to explore effective design processes and activities for engaging clinical domain stakeholders in AI concept ideation. Specifically, we utilize a design ideation resource, namely \textit{the AI Brainstorming Kit} \cite{yildirim2023creating}, that we developed in our prior work to explore \textit{how} to navigate early phase AI innovation within a multidisciplinary team.}

\vspace{-5mm}

\subsection{\textcolor{black}{Designing AI for Healthcare}}
\textcolor{black}{Healthcare is a complex product-service ecosystem consisting of many stakeholders (e.g., clinicians, patients, healthcare managers, insurance providers, regulators, etc) \cite{noortman2022breaking, kleinsmann2020new, thieme2023foundation}. A large body of research has explored the iterative design of healthcare products and services with a focus on stakeholder engagement in the early design stages \cite{huynh2022design, reay2021initiate, cunningham2019co, berry2019supporting, robertson2020if, zajkac2023clinician}. In recent years, the advances in AI and the availability of high-density datasets, such as patient electronic health records (EHR), have enabled a new wave of innovations, spanning systems that support diagnosis, treatment recommendations, and automated documentation. However, similar to other domains, AI systems in healthcare have a poor track record; they largely fail when moving from research labs to clinical practice \cite{jesso2022inclusion, yildirim2021technical, seneviratne2020bridging, yang2016investigating, west2020design, dexter2011health, craig2020context}. The clinical utility of these systems remain often unclear \cite{ghosh2023framing, galsgaard2022artificial, thieme2023designing}; as a result, clinicians often do not use them \cite{yang2016investigating, thieme2020machine}. Recent HCI research has developed healthcare AI systems with special attention to challenges around workflow integration \cite{beede2020human, burgess2023healthcare}, calibrating clinician trust \cite{jacobs2021designing, yang2023harnessing, sivaraman2023ignore}, transparency and setting mental models \cite{hirsch2017designing, cai2019hello}, and risks of biases and harm \cite{wilcox2023ai}. Relatively little work engaged healthcare stakeholders in the early stages of AI development to envision concepts that leverage AI capabilities or explore data requirements with an eye for downstream applications \cite{noortman2022breaking, yang2023harnessing}. Our work aims to address this gap, specifically within the context of intensive care.}

\textcolor{black}{The intensive care unit (ICU) is a complex, team-based healthcare setting involving many clinicians (e.g., attending physicians, fellows, residents, nurse practitioners, respiratory therapists) providing round-the-clock care for critically ill patients \cite{reddy2001coordinating}. Prior HCI work on ICUs focused on conducting field studies to understand clinician needs and workflows \cite{reddy2001coordinating, reddy2006temporality, kaltenhauser2020you}, and developing technical systems and interventions (e.g., automating patient note documentation \cite{wilcox2010physician, gopinath2020fast}, reducing alert fatigue and interruptions  \cite{srinivas2016designing, cobus2018vibrotactile, cabral2019beepless}). AI research advances in ICU demonstrate systems that predict treatment medications \cite{suresh2017clinical}, predict if a patient will need a ventilator \cite{suresh2017clinical}, predict patient discharge and readmission \cite{lin2019analysis}, and predict the onset of conditions like sepsis \cite{nemati2018interpretable}. While these proof-of-concept models indicate an initial feasibility, it remains unclear whether clinicians need help with these tasks. A recent study interviewed ICU physicians and nurses to elicit \textit{what predictions would be useful} \cite{eini2022tell} and found that clinicians desire predictions around patient trajectory and prioritization, mainly to reduce the high cognitive load rather than help with decision making. We build on this line of work to explore data and AI as design materials for ICU to identify clinically relevant and feasible AI use cases.}

\vspace{-5mm}

\section{Overview of Design Process}

We wanted to develop more effective approaches to multidisciplinary brainstorming of AI concepts, especially in the early phases of ideation and problem formulation. Building on prior literature that noted successful AI innovation teams ideate before selecting what to build, we set out to tackle the challenge of ideation within a project that focused on AI innovation in the ICU.

Our academic research team (n=22) included 6 HCI, 6 data science, and 10 healthcare experts. The HCI researchers had backgrounds in interaction design, service design, and data visualization; they brought expertise in human-AI interaction and ideation. The data science members had backgrounds in data analytics, healthcare analytics, and AI research; they brought expertise in AI capabilities and what could be built with the dataset. The healthcare members all had experience in critical care medicine and included 4 attending physicians, 2 fellows, 2 nurses, and 2 non-clinical healthcare experts. They brought expertise in clinician needs. Table~\ref{tab:participants} provides a summary of our teams’ composition.

We engaged in an iterative, reflective design process \cite{buxton2010sketching, zimmerman2007research, reis2011lean} to explore AI opportunities for the ICU, particularly to search for use cases that leveraged our ICU dataset. We conducted a three-phase study. The first phase focused on brainstorming; we conducted two ideation workshops within our team to identify clinically relevant and buildable use cases.
The second phase focused on problem formulation; we conducted a design workshop to detail a subset of 12 concepts. The third phase focused on sketching and co-design; we created low fidelity sketches for an AI concept we had generated. We conducted four co-design sessions with 11 clinicians who had not been involved in our study to elicit feedback on the design concept. Below, we provide a brief overview of the ICU dataset our team had access to. We then present each phase in subsequent sections, unpacking the research goals, design activities, and insights gained.

\begin{table}
  \caption{Our team consists of data science and AI researchers (DS), clinicians and healthcare experts (H), and human-computer interaction researchers (HCD).}
  \label{tab:participants}
  \centering
  \begin{tabular}{lllllll}
    \toprule
    ID&W1&W2&W3&Role&Exp.&Gn.\\
    \midrule
    DS1 & \checkmark &   & \checkmark & Data Scientist & 10+yrs & F \\
    DS2 & \checkmark & \checkmark & \checkmark & Data Scientist & 3-5 yrs & M \\
    DS3 & \checkmark & \checkmark & \checkmark & Data Analyst & 5-7 yrs & M \\
    DS4 & \checkmark & \checkmark &  & Healthcare Analyst & 10+yrs & M \\
    DS5 & \checkmark &  & \checkmark & AI Researcher & 5-7 yrs & M \\
    DS6 & \checkmark &  &  & AI Researcher & 5-7 yrs & F \\
    H1 & \checkmark & \checkmark &  & ICU Physician & 10+ yrs & M \\
    H2 &  &  & \checkmark & ICU Physician & 10+yrs & F \\
    H3 &  & \checkmark & \checkmark & ICU Physician & 10+yrs & F \\
    H4 & \checkmark & \checkmark & \checkmark & ICU Physician & 10+yrs & M \\
    H5 &  & \checkmark &  & Critical Care Fellow & 5-7 yrs & F \\
    H6 &  & \checkmark &  & Critical Care Fellow & 5-7 yrs & M \\
    H7 & \checkmark & \checkmark & \checkmark & Nurse Practitioner & 5-7 yrs & F \\
    H8 &  & \checkmark & \checkmark & Nurse Practitioner & 5-7 yrs & F \\
    H9 & \checkmark & \checkmark & \checkmark & Healthcare expert & 10+ yrs & F \\
    H10 & \checkmark &  &  & Healthcare expert & 10+ yrs & M \\
    HCD1 & \checkmark & & \checkmark & HCI/AI Researcher & 10+yrs & M \\
    HCD2 & \checkmark & \checkmark &  & HCI/AI Researcher & 3-5 yrs & M \\
    HCD3 & \checkmark & \checkmark & \checkmark & HCI Researcher & 10+yrs & M \\
    HCD4 & \checkmark & \checkmark & \checkmark & HCI Researcher & 5-7 yrs & F \\
    HCD5 & \checkmark & \checkmark & \checkmark & Service designer & 5-7 yrs & F \\
    HCD6 & \checkmark & \checkmark & \checkmark & Service designer & 5-7 yrs & F \\
    \bottomrule
\end{tabular}
\end{table}

% Double column figure
\begin{figure*}
\centering
  \includegraphics[width=2.1\columnwidth]{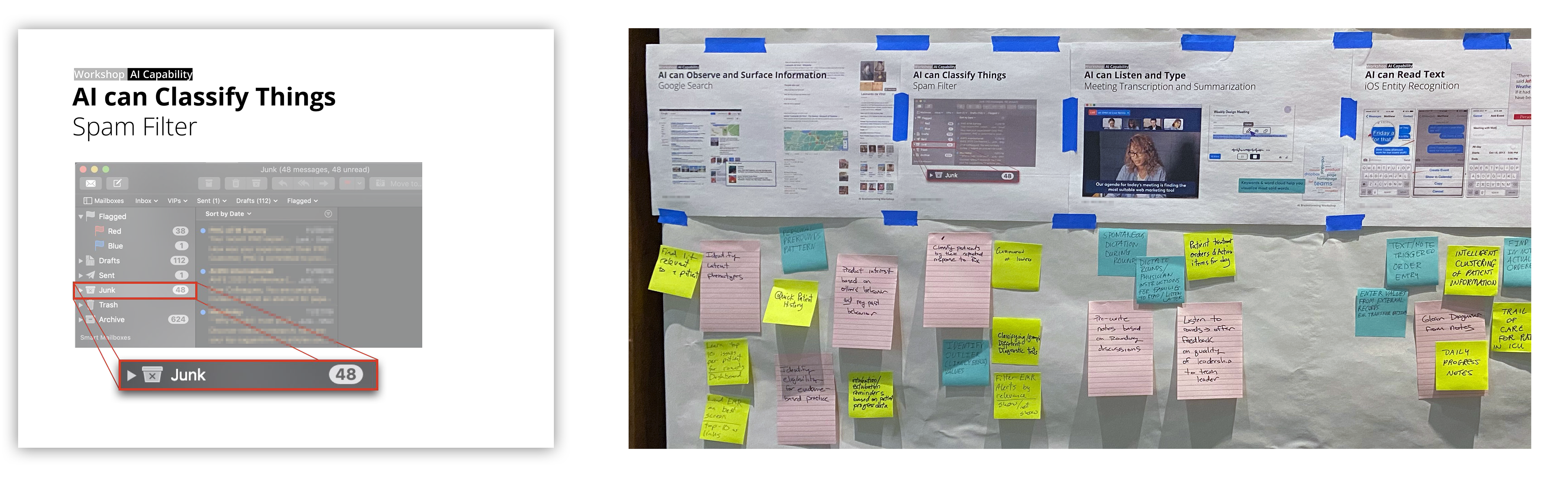}
  \caption{\label{fig:capability} An AI capability abstraction and example (left), poster printouts to prompt ideation across each capability (right).
  \Description{On the left, a poster with text "AI can classify things: spam filter" and image of a spam filter interface. On the right, printout posters with sticky notes below them.}
}
\end{figure*}

\subsection{The ICU Dataset}

The objective of our project was to broadly explore how our ICU dataset might be used to improve critical care medicine. Data availability is crucial for enabling AI capabilities \cite{yang2020re}. However, prior studies on envisioning future AI solutions often do not draw from a particular dataset, and instead focus on what would be possible with pretrained models or data that could be collected \cite{yang2019sketching, morrison2017imagining, liao2023designerly}. While there is research exploring real-world datasets with domain experts, these studies often do not focus on AI innovation or technical feasibility of the envisioned systems \cite{dove2014using, bogers2016connected}. Our focus was on bounding ideation with a real world data set to address this gap.

Our dataset consisted of two parts: electronic healthcare records (EHR) and staffing metadata. Similar to the publicly available MIMIC dataset ~\cite{johnson2019mimic}, the EHR data included patient level variables, such as hospitalization (e.g., age, gender, race, discharge disposition, admission and discharge dates, etc.); diagnosis and procedure codes, comorbidities; medications; clinical events, mechanical ventilation; and others with a total of 15 variables. The staffing metadata included the transformation of patient level variables to anonymously identify the unique care providers across different roles (i.e., physicians, nurses, respiratory therapists) who provided primary care for each patient at a shift-level. The creation of this additional dataset was motivated by prior literature that indicated whether and how long individual care providers had worked together in the same team impacts the quality of care in the ICU \cite{dietz2014systematic}. The dataset was collected across 39 ICUs from 18 hospitals on the East Coast of the United States between 2018 and 2020 (see supplementary materials for a high-level overview of the data schema).

\section{Phase 1: Brainstorming AI Concepts}
We wanted to explore how we can effectively brainstorm AI concepts as a multidisciplinary team. The healthcare members would bring expertise on what is relevant and what might transform critical care practice. The data science members would bring expertise in what might be possible to build. The HCI members would bring expertise in ideation. Our goal was to rapidly and broadly explore the problem-solution space to identify clinically relevant and buildable AI concepts to improve intensive care medicine.
\textcolor{black}{Our prior research presented an overview of this initial phase in the context of the development and assessment of \textit{the AI Brainstorming Kit} – a resource that captured AI capabilities and real-world examples to scaffold cross-disciplinary ideation \cite{yildirim2023creating}. In this work, we provide a detailed account of the methodology and elaborate on workshop facilitation, selection of AI examples, and concept assessment and prioritization.}

\vspace{-2mm}

\subsection{Method}
We chose to conduct design workshops, a commonly used method in design-driven innovation \cite{dove2014using, reis2011lean}. We conducted two workshops within our team. Each workshop had 15-17 participants involving at least one participant from each role (i.e., physician, nurse, healthcare expert, data scientist, HCI researcher). Table~\ref{tab:participants} details the involvement of participants in each workshop session. Workshops were sequential such that the outcome of a workshop informed the goals and activities of the following workshop.

A part of the challenge was the preparation and structuring of the brainstorming activities. Below, we present our thinking behind each workshop, along with details on the set of activities.

\subsubsection{Workshop 1: User-centered approach}
Our first workshop followed a traditional user-centered approach. In preparation for the workshop, we had informal discussions to elicit the domain expertise of our healthcare team members. We discussed pain points and potential themes for brainstorming, both based on lived experiences and our expertise working in healthcare innovation. These preparations resulted in ``how might we'' prompts that we used to drive ideation (e.g., \textit{How might we help clinicians in orchestrating a sequence of tasks? How might we support the adoption of evidence-based practice? How might we reduce clinicians’ burden with documentation tasks?}). Inspired by design thinking methods \cite{ideo2009design}, we set our objectives as `thinking outside the box’ and `deferring judgment’ to let go of thinking about the limits of technology.

We conducted a 2-hour in-person workshop. The workshop agenda included the introduction of goals (10 min), two consecutive ideation sessions with a short break in between (30 min), impact-effort assessment of concepts (30 min), and a short debriefing and reflection (10 min). During the ideation sessions, each team member reviewed the how-might-we prompts to first ideate individually. They next shared concepts within the group to brainstorm collectively. We used large papers, sticky notes, and markers to note down concepts. At the end of the session, we selected a subset of concepts based on the team’s interest, and placed these on a large impact-effort matrix \cite{nngroup2018impact} by getting group consensus on whether the concept was relevant and useful to critical care (impact) and if it would be easy or difficult to implement (effort). Following the workshop, the lead HCI researcher further analyzed concepts to assess the coverage of design space (see section \ref{analysis}).

\subsubsection{Workshop 2: User-centered and tech-centered approach}
Following the first workshop, we had concerns that our concepts mostly focused on places where near-perfect AI performance was needed for the use cases to be valuable – a well-documented pitfall in AI design literature \cite{yang2020re, dove2017ux, feng2023ux, subramonyam2022solving}. Building on recent research \cite{yildirim2023investigating}, we decided to bring elements from the matchmaking method \cite{bly1999design} to blend user-centered thinking and tech capability-driven approaches. Prior to the workshop, we selected a subset of AI capabilities and examples \textcolor{black}{from the AI Brainstorming Kit \cite{yildirim2023creating}}. Hoping to move away from envisioning use cases that required high AI accuracy or performance, we mostly selected examples where moderate performance and imperfect AI capabilities produced value.

%Table double column
\begin{table*}
  \centering
  \includegraphics[width=0.7\paperwidth]{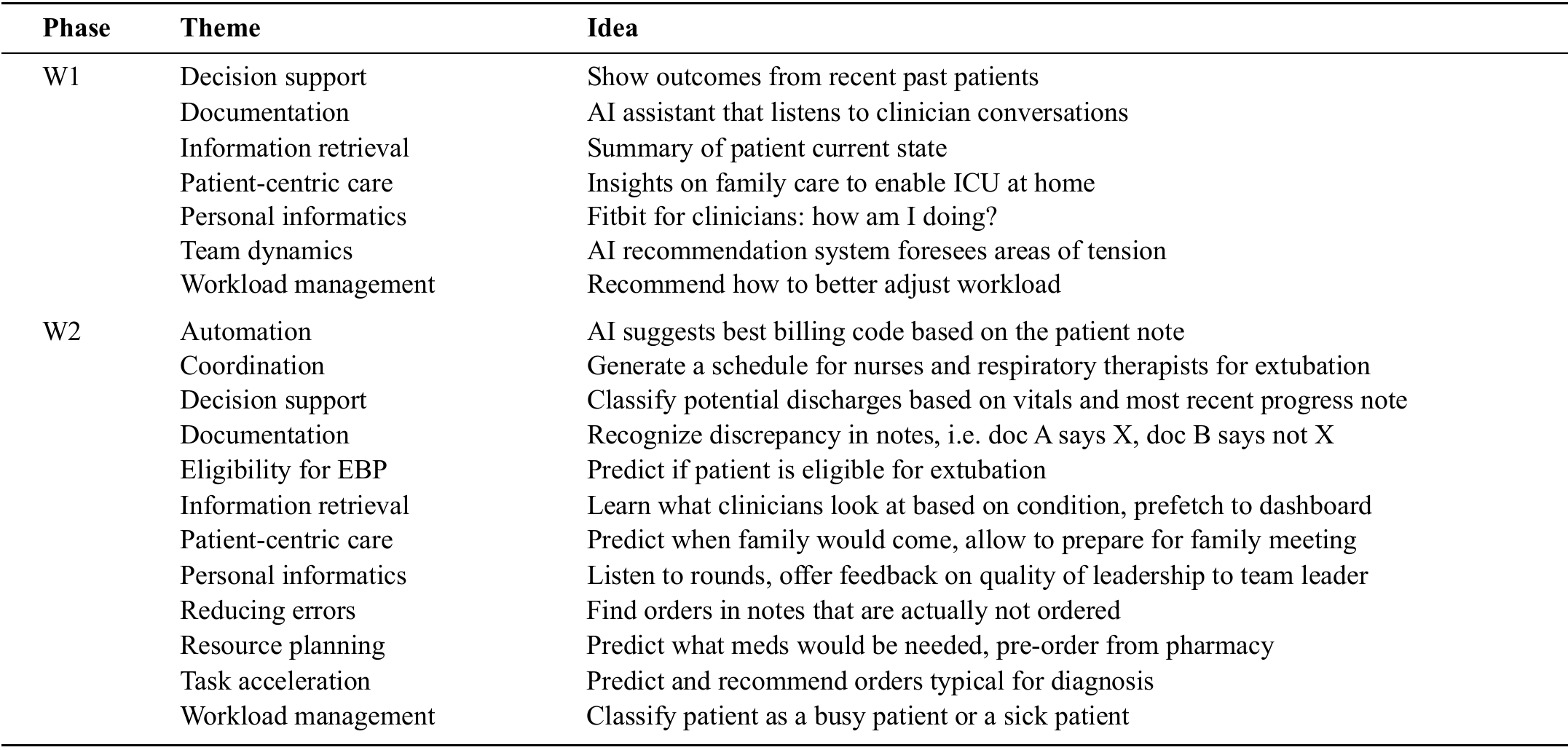}
  \setlength{\abovecaptionskip}{-8pt}
  \caption{\label{tab:ideas}High level themes and example concepts from first and second ideation workshops.}
\end{table*}

The capabilities and examples included \textit{observe and surface information} (contextual web search); \textit{classify things} (spam filter); \textit{listen and type} (real-time meeting transcription); \textit{read text} (text message entity recognition); \textit{predict text} (email sentence completion); \textit{cluster similarities} (online shopping recommender system); \textit{discover patterns} (smartwatch activity trends) [see Figure~\ref{fig:capability} and supplementary materials]. Selection and curation of capabilities were not meant to be exhaustive; similar to prior work \cite{morrison2017imagining, yang2019sketching, yildirim2022experienced}, our goal was to have a \textit{good enough} subset to inspire ideation.

We conducted a 2-hour in-person workshop following the same structure as in the first workshop. However, this time we started by reviewing the AI capabilities and examples we had prepared in the form of slides during the introduction session (10 min). We used the slides as poster printouts to prompt ideation across each specific capability. For instance, talking about ``email spam filter'' as an example of binary classification (spam or not spam), we probed if we could envision use cases where classifying things as important or not important, or as urgent or not urgent could be useful. Ideation sessions were followed by impact-effort assessment and debriefing, as in the initial workshop.

\subsubsection{Data Collection and Analysis} \label{analysis}
\begin{figure*}
  \centering
  \includegraphics[width=1.8\columnwidth]{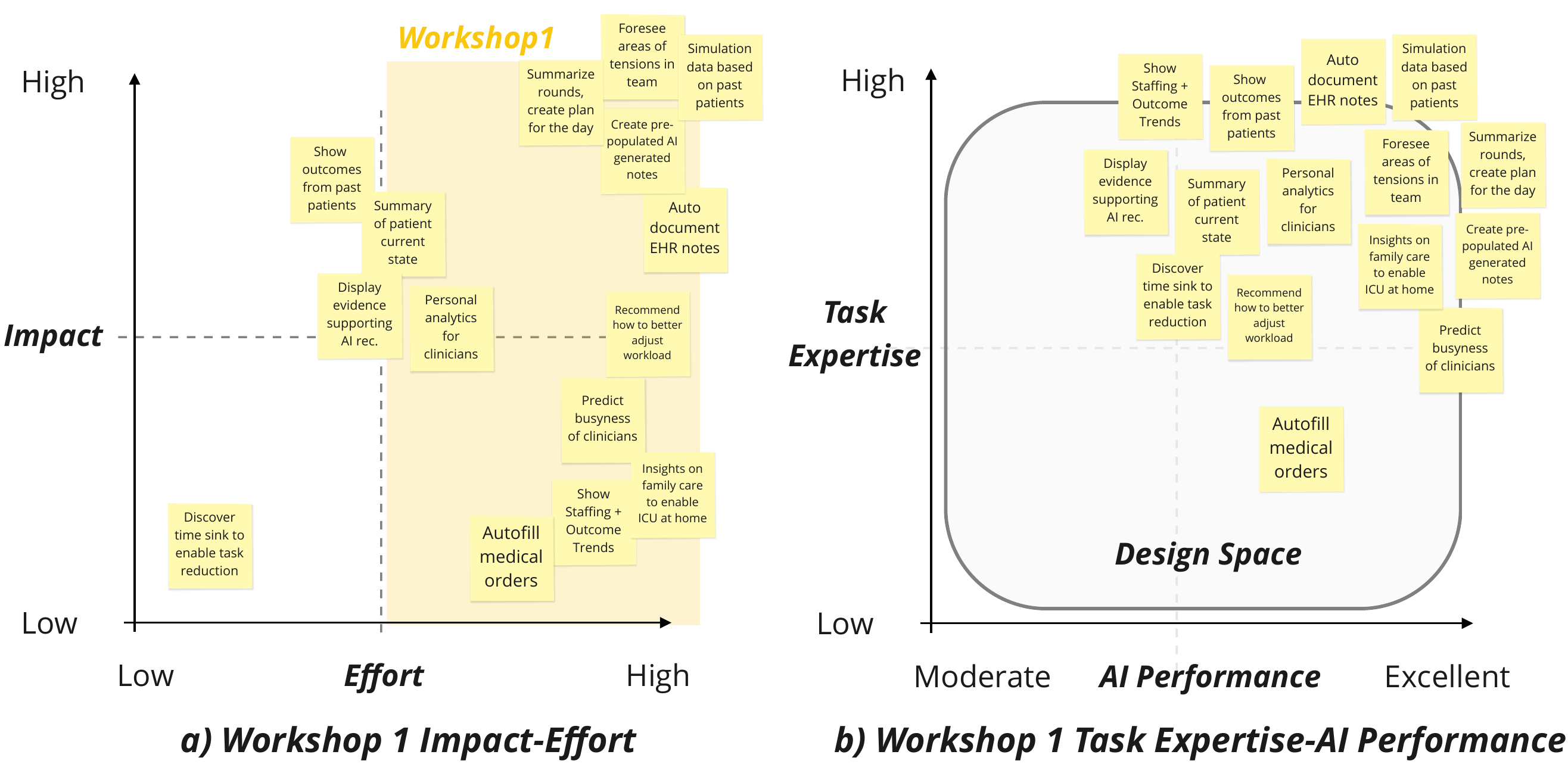}
  \setlength{\belowcaptionskip}{-5mm}
  \caption{\label{fig:workshop1} Our first workshop resulted in ideas that were technically difficult, some of which were clinically relevant.
  \Description{On the left, a two by two impact-effort matrix showing several yellow sticky notes on the upper and lower right quadrant. On the right, a task expertise-AI performance matrix showing several yellow sticky notes that are on the upper right quadrant.}
}
\end{figure*}

Workshops were audio and video recorded, and transcribed. The analysis included reviewing (1) the transcripts using interpretation sessions, and (2) workshop outcomes using affinity diagramming \cite{karen2017contextual, martin2012universal}, and the task expertise-model performance matrix \cite{yildirim2023creating} – a new assessment tool our team had created to assess the breadth of AI problem-solution space (see Figure \ref{fig:workshop1}b). This matrix broke down concepts into a two-by-two matrix based on two dimensions: \textbf{task expertise} (\textit{how much human expertise or intelligence does this task require?}) and \textbf{model performance} (\textit{what is the minimum quality needed for users to experience AI as useful?}). The analysis focused on identifying key themes for the concepts, challenges in collaboration, and the impact of design activities on workshop outputs. Two authors led the analysis, before sharing the results and insights with the entire study team for further review and discussion. We then iteratively discussed and restructured the emerging themes to seek agreement on interpretations across members.

% Double column figure

\subsection{Findings}

In this section we present workshop results by describing (1) outcomes detailing the quality of concepts, and (2) our reflections on what worked, what did not work, and what was unexpected.

\subsubsection{Workshop 1 Outcome}
The first workshop was effective at getting all members of the team to ideate. However, the outcomes seemed to cover a narrow space. Our impact-effort assessment showed that the majority of our concepts were difficult to build, while only about half seemed relevant and useful for critical care medicine (Figure~\ref{fig:workshop1}a). Our analysis of high-level brainstorming themes also indicated a lack of breadth: more than a third of concepts focused on clinical decision making, and another third described systems that automated documentation. A few of the concepts described new AI-enabled interactions. One concept described a system that forecasts expert disagreement. For example, it might predict that a nurse would not perform a specific assessment because they viewed the patient as not qualifying while the physician would want the assessment to have been performed. Another described an AI assistant that listens and transcribes conversations between clinicians.

Overall, our team collectively felt that the concepts were not very novel. Most of the concepts addressed existing interactions instead of proposing new ways of working. Concepts often described latent desires around trust, feedback, and explainability (e.g. \textit{AI can take feedback on why it is wrong}); human-AI interaction forms (e.g. checklist, chatbot, recommendation system, conversational assistant); desired system behaviors (e.g. \textit{recommendation is not intrusive, recommendation comes when ICU team is together}); and pain points (e.g. \textit{placing orders is a burden; I want to eliminate and delegate tasks}).

Similar to the impact-effort assessment results, our task expertise-AI performance analysis showed that most of the concepts mapped to the upper right region (high expertise-excellent performance), missing the larger design space (Figure~\ref{fig:workshop1}b). Concepts often required near-perfect AI performance or accuracy to be useful. For instance, anticipating clinician disagreement or predicting if a nurse will not perform an assessment can be useful \textit{only} if the AI system can correctly capture 9 cases out of 10. The system would not be useful if it incorrectly flags situations or can only catch cases correctly once in a while. Our concepts also seemed too focused on situations with high uncertainty where the task is difficult even for highly trained experts (e.g., clinical decision making, anticipating potential disagreements).

\textbf{Post-workshop reflection.} Our brainstorming workshop was successful in that our multidisciplinary \textcolor{black}{team} generated many concepts for potential AI use cases.
\textcolor{black}{Data science and healthcare team members found the brainstorming exercise novel, as they had not previously engaged in formal, structured brainstorming or human-centered design perspectives.
However, assessment of the workshop outcomes showed that} the concepts were not of the quality we wanted. Our process was not generating any concepts that were easy to develop; \textit{low hanging fruit} where moderate AI performance could generate value in the ICU. Some concepts did not require AI, and several called for data that does not exist. Reflecting on the outcomes, we set a new goal to move ideation towards \textit{situations where moderate AI performance could still generate value}.

% Single column figure
% \begin{figure*}
%   \centering
%   \includegraphics[width=1\columnwidth]{figures/workshop1.pdf}
% %   \setlength{\belowcaptionskip}{-5mm}
%   \caption{\label{fig:workshop1} Our first workshop resulted in concepts that were technically difficult, some of which were clinically relevant (a). We realized that most concepts required near-perfect AI performance to be useful, missing places where moderate performance AI could be useful (b).
%   \Description{On the left, a two by two impact-effort matrix showing several sticky note concepts on the upper right and lower right quadrants. On the right, a two by two task expertise-AI performance matrix showing sticky notes on the upper right region with high expertise-excellent performance concepts.}
% }
% \end{figure*}

% % Double column
\begin{figure*}
  \centering
  \includegraphics[width=1.8\columnwidth]{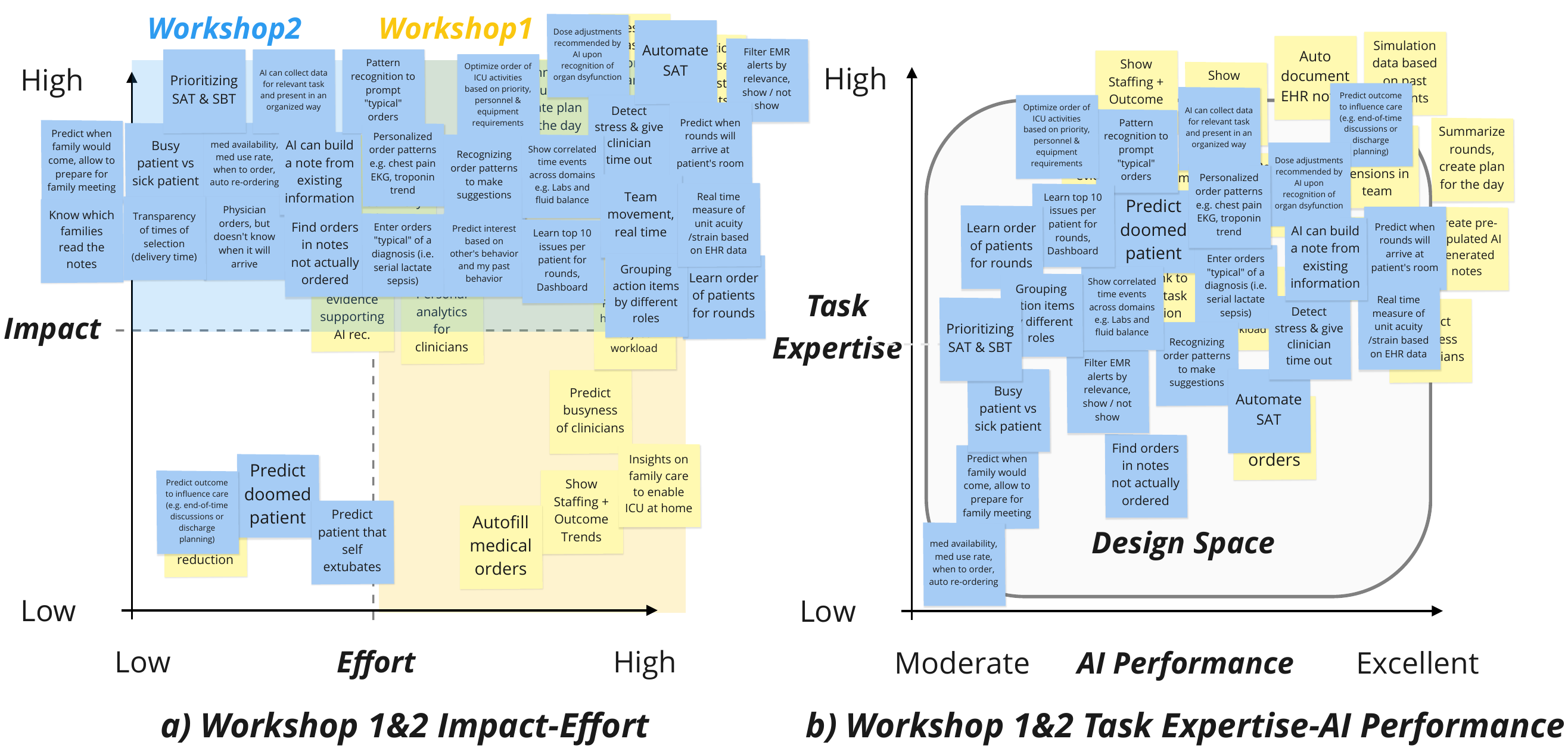}
  \setlength{\belowcaptionskip}{-5mm}
  \caption{\label{fig:workshop2} In the second workshop, concepts moved towards (a) low-effort and high-impact; (b) from high expertise-excellent performance to medium expertise-moderate performance.
  \Description{On the left, a two by two impact-effort matrix showing several blue sticky notes on the upper right and upper left quadrants overlaid on top of yellow sticky notes from the first workshop that are on the upper and lower right quadrants. On the right, a task expertise-AI performance matrix showing several blue sticky notes towards covering the whole space overlaid on top of yellow sticky notes from the first workshop that are on the upper right quadrant.}
}
\end{figure*}

\vspace{-2mm}

\subsubsection{Workshop 2 Outcome}
The second workshop led to concepts that mapped to a broader set of themes. This was one type of concept quality we were particularly focused on. Examples included AI systems that would improve coordination between clinicians (e.g. \textit{generate a schedule for nurses and respiratory therapists for extubation}); systems that improved logistics and resource allocation (e.g. \textit{predict which medications would be needed based on current patients and pre-order from pharmacy}); systems that inferred workload and effort, possibly in support of dynamic staffing (e.g. \textit{classify patients as sick or busy}); systems that better support attention management (e.g. \textit{classify alerts as urgent or not urgent}); systems that improve efficiency, particularly around data entry and documentation (e.g. \textit{predict and recommend orders typical for diagnosis}); systems that anticipate needed information (e.g. \textit{learn relevant information based on patient conditions}).

% Single column figure
% \begin{figure*}
% \centering
%   \includegraphics[width=1\columnwidth]{figures/capability1.png}
%   \caption{\label{fig:capability} An AI capability abstraction and example (left), poster printouts to prompt ideation across each capability (right).
%   \Description{On the left, a poster with the text "AI can classify things: spam filter" and an image of a spam filter interface. On the right, printout posters with sticky notes below them.}
% }
% \end{figure*}

% Single column
% \begin{figure*}
%   \centering
%   \includegraphics[width=1\columnwidth]{figures/workshop2.pdf}
%   \setlength{\belowcaptionskip}{-5mm}
%   \caption{\label{fig:workshop2} In the second workshop, (a) our concepts moved towards the upper left quadrant on impact-effort matrix: we were able to identify low effort-high impact concepts; (b) cover a larger design space on the task expertise-AI performance matrix with medium expertise-moderate performance concepts.
%   \Description{On the left, a two by two impact-effort matrix showing several sticky note concepts on the upper right and upper left quadrant. On the right, a two by two task expertise-AI performance matrix showing many sticky notes covering a large design space.}
% }
% \end{figure*}

In addition to these new themes, we generated concepts that built on the themes from the previous workshop, including decision support (e.g. \textit{predict if the patient is eligible for extubation}); documentation (e.g. \textit{generate a draft patient note based on available information}), and automation of menial tasks (e.g. \textit{recommend best billing code based on the patient note}). Table~\ref{tab:ideas} lists the high level themes and example concepts from each round of workshops.

Using AI capabilities and examples served as a springboard for our team to recognize situations where a capability could be useful to then effectively transfer that capability to a use case. For example, a nurse practitioner envisioned classifying patients into two groups, sick patients and busy patients. This mirrored the \textit{classify things} capability. Sick patients typically require more attention. Busy patients included patients who needed many time-consuming procedures: \textit{``Is this a busy patient? Or is this a sick patient? It would be useful for managing nursing tasks to tell the difference between a patient who's incredibly sick, but doesn't have a lot of tasks. … [versus] if they have a lot of weeping wounds or something like that, that can make for a very busy patient.'' (Nurse 2, H8)} This concept hinted at the potential for more dynamic staffing or could be used to \textcolor{black}{balance work difficulty and staff expertise} across an ICU. Another capability, \textit{observing and surfacing information}, spurred the concept of learning what EHR screens and information clinicians looked at based on patient condition in order to prefetch or highlight \textcolor{black}{relevant patient history information} on a dashboard.

In impact-effort assessment, our concepts moved towards the upper left quadrant: we were able to identify concepts that required low implementation effort with potentially high-impact (Figure~\ref{fig:workshop2}a). The task expertise-model performance assessment also revealed that the concepts moved from high expertise-excellent performance to medium expertise-moderate performance (Figure~\ref{fig:workshop2}b). For example, generating an ordered list of patients for rounds based on the uncertainty of what to do seemed relatively low-risk. A moderate quality, draft triage list is still better than no prioritization; the clinical team will still attend to all the patients in the ICU. Interestingly, in expanding the solution space towards situations where moderate AI performance could be useful, we moved beyond high-stakes situations with great uncertainty (e.g., clinical decision making) and produced concepts for relatively underexplored places (e.g., coordination, managing workload, anticipatory information retrieval).

\subsubsection{Post-study reflection.}
Discussing specific AI capabilities and examples prior to the workshop seemed to have a significant impact on the outcomes of ideation, yielding a broader design space where a mediocre, imperfect AI model would still provide enough value for clinicians. We also noticed that explicitly talking about AI capabilities provided our team with a shared language. Unlike the first round, most sticky notes described interaction concepts starting with capability verbs (e.g. \textit{detect, recognize, classify, notice, predict, generate}...). Using this language, clinicians probed data scientists about technical possibilities. \textit{``Can AI notice the sequence of orders? ... Can AI cluster tasks?''} Ideation became a collective conversation to discuss what would be doable, how that would produce value for users, and whether any relevant data was captured.

Although the quality of the concepts improved, we still encountered challenges. First, our assessment showed that while our concepts were grounded in what's technically possible, only a few of them were implementable using our specific ICU dataset. Most concepts required additional data collection or instrumentation (e.g. tracking clinician clicks in UI to learn and pre-fetch information to dashboards). In some cases, the data existed but it was not in our dataset (e.g. unstructured text from clinical notes), rendering our concepts infeasible unless we sought out more and different data. Overall, the ideation exercise was valuable for informing data collection for future implementations, but we were ignoring the value of our ICU dataset in our ideation. We needed concepts we could build using our data to create immediate value for clinicians.

We also noticed that similar to other healthcare innovation research \cite{yang2023harnessing}, we had a tendency to attribute familiar interaction forms, such as alerts, to specific capabilities and concepts based on past experiences. For instance, while we liked the concept of classifying patients, we always seemed to imagine this as an alert or a reminder. Given the well-known research on alert fatigue and clinician burnout \cite{cash2009alert}, this seemed problematic. Our fixation on existing forms bound to a capability posed a threat to ideation, as the team would dismiss concepts based on known problems with the familiar forms. As prior research reported \cite{yang2019sketching}, we found ourselves trying to separate the inference (e.g., predicting that a patient would need a scan) from the interaction (e.g., recommending the action to a clinician or proactively ordering a scan).

Relatedly, rapid ideation resulted in surface level concepts that require further exploration. For instance, clinicians liked the concept of having a ranked list of patients to visit during rounding. However, the criteria needed to prioritize patients was not clearly defined: should it be based on sickness level (see sickest patients first) or patient uncertainty (patients where it was least obvious what to do)? In order to more effectively assess the concepts and select candidates for development, we needed more detail on what the concept was and how it might work in terms of data requirements and the form of the AI output clinicians would encounter.

\textcolor{black}{Finally, following the second workshop, discussions on how to move forward surfaced confusions and a need for increased communication within the team. While the HCI team perceived the second workshop as a success –especially from a methodological point of view– the shift in the quality of ideation was not obvious to the rest of the team. Conversely, the data science and healthcare team members found the exercise to be repetitive. The clinical team lead expressed confusion over the activities in the second workshop, probing the reasoning behind generating concepts from scratch instead of building on the existing ideas from the first workshop. To resolve concerns, the lead HCI researcher presented the post-workshop assessment of concepts, clarifying how the quality and breadth of ideation has shifted. The team then reached a consensus that the next best step would be to select a subset of ideas that could be grounded within our ICU dataset for further detailing and assessment.}

% % Double column figure
% \begin{figure*}
% \centering
%   \includegraphics[width=1\columnwidth]{figures/worksheet.png}
%   \caption{\label{fig:examples} XX.
%   \Description{XX.}
% }
% \end{figure*}

% % Single column
% \begin{figure}[h]
%   \centering
%   \includegraphics[width=0.5\columnwidth]{figures/matrix_revised.pdf}
%   \setlength{\belowcaptionskip}{-5mm}
%   \caption{\label{fig:matrix} \textcolor{black}{Once we identified high-impact ideas, we detailed feasibility by breaking down task expertise and acceptable AI performance. Medium expertise-moderate performance ideas seemed to have lower risk associated.}
%   \Description{Two by two matrix detailing Task Expertise against AI performance. Ideas from workshop 1 are on the upper left quadrant. Ideas from workshop 2 are in the middle.}
% }
% \end{figure}

% % Double column
% \begin{figure}[h]
%   \centering
%   \includegraphics[width=1\columnwidth]{figures/matrix.pdf}
%   \setlength{\belowcaptionskip}{-5mm}
%   \caption{\label{fig:matrix} Once we identified ideas with high impact, we detailed feasibility by breaking down AI task difficulty and acceptable AI performance. Medium difficulty-medium performance ideas seemed to have lower risk associated.
%   \Description{Two by two matrix detailing AI Task Difficulty against AI performance. Ideas from workshop 1 are on the upper left quadrant. Ideas from workshop 2 are in the middle.}
% }
% \end{figure}

\vspace{-2mm}

\section{Phase 2: Problem Formulation}

As we moved from ideation to problem formulation, we set three goals. First, we wanted to leverage the unique properties of our dataset, and ground our concepts in what we could realistically build. Second, we wanted to separate interaction form and AI inference when discussing concepts. Third, we wanted to deeply explore some of the concepts to have more mature conversations on their feasibility, desirability, and potential implications.

\subsection{Method}
We chose to conduct an additional design workshop that focused on problem formulation. Similar to the phase 1 study, we conducted a 2-hour in-person workshop for detailing a subset of 12 concepts. Below, we first describe how we prioritized and selected the subset of concepts prior to the workshop. We then detail the artifacts prepared for the workshop and the set of activities.

\subsubsection{Concept prioritization}
We had three criteria when selecting concepts for further development. First, we prioritized concepts based on data availability, choosing concepts that could be built using our ICU dataset. Second, we sought to cover a breadth of the design space, selecting concepts where moderate-to-good performance AI could produce medium-to-high value. Finally, we included concepts that matched our team’s research interests and expertise, excluding some concepts in subspecialty AI areas (e.g., natural language processing or computer vision-based concepts). The selected concepts included: anticipatory pre-ordering of medications; predicting medication time-to-delivery; generating a prioritized list of nurse assignments; identifying sick or busy patients; forecasting unit acuity; generating an ordered list of patients to see for rounds; predicting the eligibility of patients for extubation from mechanical ventilators; generating a coordinated schedule for extubation; identifying clinician workload patterns; identifying bias in clinical orders; predicting typical orders for diagnoses; and discovering the sequence of tasks.

\subsubsection{Workshop preparation}
\begin{figure}[h]
  \centering
  \includegraphics[width=1\columnwidth]{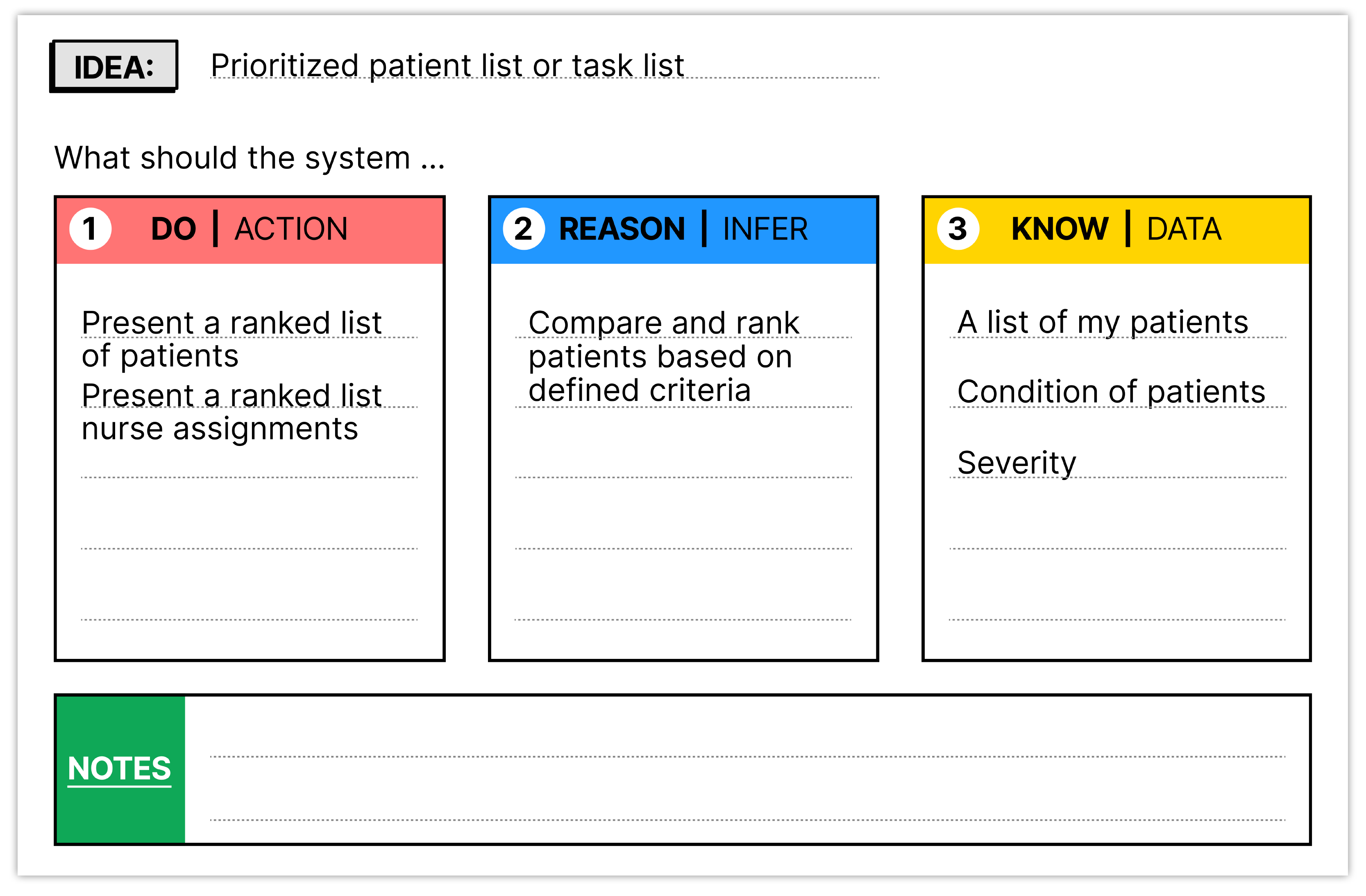}
  \setlength{\belowcaptionskip}{-5mm}
  \caption{\label{fig:worksheet} Do-Reason-Know worksheet enabled us to detail each idea in terms of model reasoning, data, and interaction form.
  \Description{A worksheet that has a top section "idea" to describe a concept, three sections in the middle "do/action", "reason/infer", "know/data", and a "notes" section at the bottom.}
}
\end{figure}

Prior to the workshop, the lead HCI researcher worked on numerous representations to untangle the inference produced by an AI model, the data needed to build the model, and the form of the AI output clinicians would encounter. Over several discussions, the team critiqued and iterated on the alternative artifacts. After rounds of iterations, we arrived at a new abstract representation: \textit{the Do-Reason-Know worksheet} (Figure~\ref{fig:worksheet}). Each section respectively captures the interaction (do), model reasoning and inference (reason), and data requirements (know).

The worksheet builds on the classical input-model-output representation commonly used in machine learning \cite{google2022ml}, yet it furthers the existing artifacts in two aspects. First, it captures both the inference and the delivery of the inference for separating the model behavior (e.g. rank patients) from the interaction behavior (e.g. present a list where critical patients are displayed at the top). Second, it balances the model-centric view with a user-centric view by flipping the starting point (end user interaction instead of AI input or output), and embedding the desired system behavior into problem formulation from the beginning. In preparation for the workshop, we pre-populated the worksheets with the concept names and any other relevant information that was discussed in prior workshops (e.g. a potential data source our team had referred to related to a particular concept).

% Single column
% \begin{figure*}
%   \centering
%   \includegraphics[width=0.5\columnwidth]{figures/worksheet.pdf}
%   \setlength{\belowcaptionskip}{-5mm}
%   \caption{\label{fig:worksheet} The Do-Reason-Know worksheet enabled us to detail each concept in terms of model reasoning, data, and interaction form.
%   \Description{A worksheet with sections "do/action", "reason/infer", "know/data".}
% }
% \end{figure*}

\subsubsection{Workshop activities}
We conducted a 2-hour in person workshop. The workshop kicked off with a short review of the worksheet and the 12 concepts we pre-selected (15 min). Then, we divided into two groups, where each group collectively discussed and detailed 6 concepts (90 min). We used worksheet printouts as a starting point and detailed each section by adding sticky notes. For instance, when deliberating on \textit{predicting whether a patient might need a certain procedure (e.g. surgery, intubation)}, we discussed if the time of a procedure is documented and whether there were relevant actions or events we could use as proxies (e.g. bleeding prior to surgery). We concluded with a brief reflection and discussion on the next steps (10 min).

\subsubsection{Data Collection and Analysis}
We audio and video recorded and transcribed the workshop. We documented the worksheet printouts, and analyzed the transcripts and artifacts using the same methods as in Phase 1 (see section~\ref{analysis}).

% % Double column

\subsection{Findings} 
We first present insights from the workshop capturing our process of problem formulation. We then reflect on the use of the Do-Reason-Know worksheet in concept detailing.

\subsubsection{Workshop Outcome}

One of our goals was to focus on low-risk, medium-value concepts. Throughout the workshop, we reworked our concepts in a way that reduced the required model performance to help us identify relatively simple, low-risk AI concepts. We repeatedly asked \textit{``Is there a simpler, dumber version of this concept that is still `good enough’ to produce value?''} Below, we share three concepts to illustrate how this approach helped us effectively formulate concepts.

\textbf{Predicting if a mechanically ventilated patient is eligible to receive a breathing trial, instead of predicting if the patient should be extubated.} Liberation from mechanical ventilation is a complex process that requires coordinated actions of nurses, respiratory therapists, and physicians. It involves two integrated actions. Typically, the nurse assigned to a specific patient will perform a \textit{Spontaneous Awakening Trial (SAT);} they will cut off a patient's sedation and observe if they can tolerate being awake. Next, the respiratory therapist, who is typically in charge of making changes to the ventilator settings, will perform a \textit{Spontaneous Breathing Trial (SBT)}. They will suspend breathing support and observe how well patients take over their own breathing. These assessments allow the team to decide if a patient can be extubated (liberated from a ventilator).

Remaining on a ventilator is associated with several adverse outcomes including delirium, pneumonia, and heart damage; however, extubating the patient and taking them off the ventilator too soon leads to another host of problems \cite{kahn2006hospital, morandi2011sedation, hickey2020mechanical}. When one of the steps gets missed (SAT and SBT), then the clinical team lacks the information to make a decision about extubation, meaning the patient remains on the ventilator for another day.

Our initial concept around patient extubation focused on \textit{predicting if a patient will successfully extubate} and \textit{discovering the right amount of sedation for a patient on a ventilator}. These are hard problems that need excellent model performance and very high quality healthcare data, which may not exist. During our discussions, clinician team members shared that physicians can become risk averse when extubations fail. They speculated that this might result in patients remaining on a ventilator longer than needed.

With this in mind, we turned our attention to the execution of SAT/SBT procedures instead of the clinical decision making for patient extubation. This led the concept towards \textit{predicting a patient's eligibility to receive SAT/SBT}. This is a comparatively low-risk, moderate-performance, and medium-value concept, as it focuses on an intermediate, safe-to-perform action rather than a critical decision.

\textbf{Predicting medication availability and anticipatory ordering.} One of the promising concepts that emerged from our ideation was predicting what medications would be needed based on the patient conditions in the unit. The concept was inspired by Amazon's anticipatory shipping \cite{spiegel2014amazon} –an AI capability and example that came up during capability-based ideation workshop– where the AI system would keep track of the inventory and pre-order medications to reduce time and uncertainty.

During problem formulation, clinician team members shared that this would be really useful for custom mixed antibiotics: \textit{``Sometimes you say `Antibiotics. Now!' and two hours later it still hasn't arrived.'' (Physician 1, H1)} They noted that delays are more likely to happen in busier wards, which can be deadly \cite{ginestra2022association}. However, clinicians were also cautious as the incorrect predictions might lead to unused medications, and therefore waste.

We broke down this concept into several lower risk concepts. First, instead of preordering, the predictions could be used only to inform the pharmacists so that they have a sense of what to expect. Second, we could instead predict time-to-medication to provide feedforward to the clinical team when placing orders. Third, a simpler approach could check for antibiotic dosing errors to prevent delays: 

\begin{quote}
    \textit{\textbf{Physician 2:} ``I want this antibiotic for my patient. When the pharmacist finally gets to it, they say, you ordered the wrong dose. Because this patient is this size, this weight and has this renal function. Something smart would be able to figure that out, like smart dosing.'' (H2)}
    
    \textit{\textbf{Data Scientist 1:} ``That's a lot easier to do. We have that history of conditions, and what was given to [patients], so maybe these kinds of predictions.'' (DS1)}

\end{quote}

% Finally, as part of bringing back our focus to low hanging fruit, we often found ourselves questioning \textit{``Do we need AI for this problem?''} In most cases, challenges and pain points stemmed from a lack of human-centered design in healthcare software systems, which could be easily addressed using rule or heuristic-based solutions.

% \begin{quote}
%     \textit{\textbf{Nurse 2:} Families can read our notes in real time, but they're [more] meaningless to them than they are to me [with acronyms and medical terms]. I would love a family-friendly note to be automatically generated. (H8)}
  
%     \textit{\textbf{HCI Researcher 3:} Something that'd be much easier, would it be good to know this family has been reading the notes? So when you're going to engage with a family you have a sense of have they been looking at it? (HCD3)}
% \end{quote}

\subsubsection{Use of the Do-Reason-Know Worksheet}
The worksheet helped to scaffold conversations around data dependency, model behavior, and interaction behavior. It allowed us to express concepts in a more refined way as we moved from sticky note concepts to more fleshed out problem formulations. It prompted us to further probe each concept in terms of how it would generate value for clinicians, and which features in our dataset could drive it, if at all possible. For instance, when discussing what \textit{patient priority} means:

\begin{quote}
    \textit{\textbf{Physician 4:} It's a two by two table. There are sick people that if you do the things you need to do, they're going to be just fine. And then there's the sick people who are uncertain. I need to pay attention to this patient in the next four hours because if I don't, six hours from now, they might be dead. … [It would be great if] it was clear who those patients were, and you didn't have to take 15 minutes to figure that out. (H4)}
    
    \textit{\textbf{HCI Researcher 1:} What information helps you determine which category that patient falls into? (HCD5)}
    
    \textit{\textbf{Physician 4:} I look at what drips they're on, what's their vent settings. You'd be looking at the amount of drip titration, certain kinds of orders, certain kinds of labs, maybe some radiology findings. I think you can observe some of that in the data. (H4)}
   
    \textit{\textbf{HCI Researcher 1:} How accurate do you feel like your rankings are after you spend fifteen minutes? (HCD1)}
    
    \textit{\textbf{Physician 2:} There can be surprises, but I'm relying on my team to give me a better idea. (H4)}
    
    \textit{\textbf{HCI Researcher 4:} Do you think it would be useful? At which point this would be most useful? (HCD4)}
    
    \textit{\textbf{Physician 2:} The idea is to reduce the cognitive load on the physician. That's probably most useful at the beginning of the day, maybe at the end of the day when we switch shifts, handing off to the other person. If there was a tool there, I might check it once or twice throughout the day like, has anything changed? (H2)}
    
    \textit{\textbf{Data Scientist 1:} Presumably in the algorithm, we could do it every four hours. (DS1)}
\end{quote}

Describing the concept with this level of detail made it clear this would function as two separate two-class classifiers. Each patient would be classified as \textit{not-sick} or \textit{sick}, and they would be classified as \textit{certain of what to do} or \textit{uncertain of what to do}. Interestingly, as the model capability and reasoning became clear, our discussions moved towards:

\begin{enumerate}
    \item \textbf{Model performance:} How accurate or robust do the predictions need to be?
    \item \textbf{Point of interaction:} When, where, and how the inference should be delivered to produce value? (e.g. \textit{are predictions available 15 minutes before or the night before?})
    \item \textbf{Risk:} What are the consequences of errors? (i.e. \textit{false positives and false negatives})
    \item \textbf{Data quality:} Is the training data trustworthy? Is it likely to introduce bias?
\end{enumerate}

Specifically, the worksheet helped with the three challenges we previously encountered. First, it allowed us to collectively define and formulate AI experiences in a way that is grounded in our dataset. Second, it allowed us to free up our concepts from existing forms by separating the interaction, AI capability, and data. Third, it informed our design deliberation and supported a deeper discussion of the concepts before starting model building and prototyping.
\textcolor{black}{For example, when discussing the concept \textit{predicting typical orders for diagnoses}, one physician likened this to a personalized contacts list in email clients, where typing upon a contact name would present the most frequent contact at the top. The personalization aspect raised some concerns: would the medication orders be based on an individual clinician's previous orders or based on a group of clinicians' orders? Physicians seemed to prefer a personalized system, which seemed more complex and costly (both in terms of model building and continuous learning). These deliberations helped us weigh cost-value tradeoffs throughout problem formulation.}

Our third workshop had an additional, unexpected benefit: our discussions helped our team to reveal existing or potential problems in our dataset. For instance, one of our ideas was around predicting patient eligibility for extubation from a mechanical ventilator to help clinicians plan for extubation. While exploring potential features in our data, we discussed whether we could use Riker scores, a numeric score for documenting the level of a patient's \textcolor{black}{sedation level} and consciousness. When discussing this concept, healthcare members shared that Riker score data were not trustworthy. The scores nurses entered into the EHR did not always reflect the actual level of \textcolor{black}{sedation}. This problematic data did not impact the quality of care as clinicians looked at the patient before making a decision. They did not make sedation decisions based on what was captured in the EHR. Thus, they never fixed this data entry problem. Interestingly, this issue is neither reported nor speculated in medical literature. Uncovering this insight early on in the process helped our team avoid using data features that clinicians did not trust.

% Double column figure
\begin{figure*}
\centering
  \includegraphics[width=2\columnwidth]{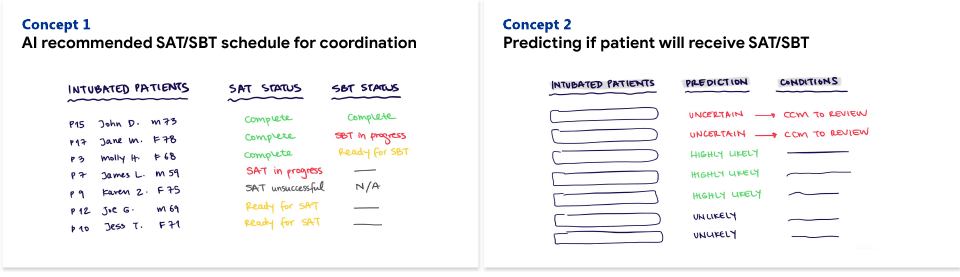}
  \setlength{\belowcaptionskip}{-5mm}
  \caption{\label{fig:sketches} Two low fidelity sketches detailing the concept of \textit{predicting if a mechanically ventilated patient is eligible to receive the SAT/SBT protocol}. We used the sketches to conduct co-design workshops to elicit feedback from nurses and respiratory therapists.
  \Description{On the left, a concept sheet capturing the concept name "AI recommended SAT/SBT schedule for coordination" and a sketch showing three columns: a list of intubated patients, SAT status, and SBT status. On the right, a concept sheet capturing the concept name "predicting if patient will receive SAT/SBT and a sketch showing three columns: a list of intubated patients, predictions indicating if patient is highly likely, unlikely or uncertain, and lastly conditions of the patients.}
}
\end{figure*}

% Single column
% \begin{figure*}
%   \centering
%   \includegraphics[width=1\columnwidth]{figures/icu-sketches1.pdf}
%   % \setlength{\belowcaptionskip}{-5mm}
%   \caption{\label{fig:sketches} Two low fidelity sketches detailing the concept of \textit{predicting if a mechanically ventilated patient is eligible to receive the SAT/SBT protocol}. We used the sketches to conduct co-design workshops to elicit feedback from nurses and respiratory therapists.
%   \Description{XX}
% }
% \end{figure*}

\subsubsection{Post-study reflection.} The problem formulation workshop with the focused worksheet activity helped us detail our concepts for further development. Following this workshop, we decided to sketch out some concepts in detail to elicit initial user feedback. \textcolor{black}{Notably, the workshop debrief revealed many insights into the felt experience of our team. For instance, the clinical team lead found the workshop series valuable from a portfolio building and de-risking point of view: \textit{`In [healthcare ML research] there is a lot of inertia towards low-risk, low-reward areas that doesn’t move the needle in a meaningful way. This exercise is really valuable because people can replicate these methods to identify lower-risk yet high-reward ideas that are worth doing. Every research portfolio should have a mix of those.’ (H1)} Reflecting on how the exercise can be improved, some clinicians shared that involving a broader set of stakeholders would be more helpful: \textit{`It might be useful to have in the room like somebody from hospital management, somebody from pharmacy … to help fill in some of the gaps, [as we have] been making some assumptions.’ (H2)} Finally, all data science team members expressed that they found the third workshop the most beneficial. It seemed to help them to gain a deeper understanding of clinical domain knowledge in relation with the data: \textit{``It's great to hear how and where the data is coming from.'' (Data Scientist, DS2}). After the workshop, several data science team members shared additional concepts or ideas on implementation details with the team based on the insights our discussions sparked.}

\textcolor{black}{From a methodological perspective, using a combination of impact-effort matrix and task expertise-AI performance matrix, along with the Do-Reason-Know worksheet allowed us to quickly sort out ideas that our team was most interested in. However, in hindsight, we noticed that dimensions, such as impact and effort, can be even more granular for a more rigorous concept assessment. For example, questions around effort included \textit{`is there any data available?’}, \textit{‘how much work is needed for data cleaning or anonymization?’}, and \textit{‘how easily can we measure and validate AI outputs?’} Moreover, the AI performance and effort (feasibility) seemed related; we repeatedly asked \textit{`what level of performance is needed?’} and based on that \textit{`how difficult is it to achieve that performance?’}. We also noted two other critical dimensions that we have not delved into: financial viability \textit{(‘how expensive is this model to build and run?'}, \textit{‘how much return on investment (ROI) is it likely to generate’)} and potential responsible AI issues \textit{(‘are there issues around privacy, fairness, data bias?’)}. We reflected that capturing these dimensions in a more nuanced manner can inform the future iterations of the Do-Know-Reason worksheet (e.g., similar to datasheets \cite{gebru2021datasheets, rostamzadeh2022healthsheet}, a comprehensive `AI concept template’ for concept proposals).}

\section{Phase 3: Sketching and Co-Design}

Following ideation and problem formulation, we chose to further develop the concept of \textit{predicting if a mechanically ventilated patient is eligible to receive the SAT/SBT protocol}. We engaged in a concurrent model development and interaction design process. The clinician and data science team members carried out the data and model work, and the HCI team members conducted co-design sessions with end users. In this section, we provide a brief overview of our sketching and co-design process to illustrate how we moved from ideation towards sketching and concept refinement, envisioning how clinicians might interact with an AI system.
% For a comprehensive account, we refer to our work detailing this case study [citation removed for blind review].

% Single column
% \begin{figure*}
%   \centering
%   \includegraphics[width=1\columnwidth]{figures/icu-sketches1.pdf}
%   % \setlength{\belowcaptionskip}{-5mm}
%   \caption{\label{fig:sketches} Two low fidelity sketches detailing the concept of \textit{predicting if a mechanically ventilated patient is eligible to receive the SAT/SBT protocol}. We used the sketches to conduct co-design workshops to elicit feedback from nurses and respiratory therapists.
%   \Description{XX}
% }
% \end{figure*}

\subsection{Method}

\subsubsection{Concept Sketching}
We created two low fidelity concept sketches detailing a shared dashboard for nurses and respiratory therapists (RTs) to support them in executing the SAT/SBT protocol for mechanically ventilated patients. The first concept displayed a dashboard with an AI-generated SAT/SBT patient schedule for better coordination (Figure \ref{fig:sketches}a). The second concept displayed a dashboard that predicted if a patient will receive an SAT/SBT based on past data, and ranked the patients based on uncertainty. In this concept, high uncertainty patients were displayed on top, so that the care team could resolve uncertainties at the beginning of the morning shift (Figure \ref{fig:sketches}b).

\subsubsection{Co-Design Workshops}
We conducted four co-design workshops with 6 RTs and 5 nurses. In each session, we had at least one nurse and one RT participant. We recruited participants through a mix of purposive and snowball sampling~\cite{heckathorn2011comment}, first reaching out to our contacts at collaborating hospitals, then expanding this set by asking participants to share relevant contacts. Workshops were conducted in-person, and facilitated by the lead HCI researcher. We first probed participants about their current practices for executing the SAT/SBT protocol. We then shared the concepts as print outs, asking them to reflect whether and how these could be useful. We provided markers and pens for participants to directly edit and comment on the concepts. Each session lasted approximately 2 hours. Participants were compensated \$250 for their time. The study was approved by our Institutional Review Board.

\subsubsection{Data Collection and Analysis}
All workshops were audio and video recorded, and transcribed verbatim. We also documented the printouts that recorded participants’ notes. We analyzed the data using the same methods described in Phase 1. 

\subsection{Findings}
\subsubsection{Workshop Outcome}
Overall, our participants perceived the concepts as valuable. They reflected that having a shared dashboard that pre-assessed a patient’s eligibility for the protocol and documented any contraindications would help them plan for patients. Several participants desired the system to not only show the patient eligibility, but also the longitudinal SAT/SBT history: \textit{`Specifically, did they meet the criteria? How long were they on it? What was the contraindication? Was this SBT done? What were the settings? Was it successful? If not why?'}
They also indicated that the system could offer meaningful category labels to indicate why a patient was categorized as ineligible: \textit{`A good category [for ineligible patients] would be seeing condition A, if they were called for an emergent reason [such as] airway protection, drug overdose.’ (RT1)}

Both nurses and RTs reflected that patients who have high uncertainty are often deprioritized as the uncertainties tend to go unresolved, resulting in eligible patients not receiving the protocol. An RT reflected that flagging these patients would be useful for the care team to review: \textit{`If the nurse charts that their neuro function is not normal, it's probably uncertain to me, the doctor needs to review. So if those were put in the algorithms and sorted out, I can tell who I'm going to see first.’ (RT5)}. However, some participants indicated that they would not trust an algorithm-based patient prioritization. They expressed a desire for the involvement of the physician, who could review this draft list to adjust the patient priority based on their goals. Finally, participants expressed that knowing high risk patients –patients who are most likely to fail the breathing trial– might be useful for planning and coordination: \textit{`If you know every one of your patients [who] is going to be absolutely terrible when you SBT them, you might want to do all your other SBTs first, and then go to get them last to make sure your nurse is with you in the room.’ (RT3)}

\subsubsection{\textcolor{black}{Post-study reflection.}}
\textcolor{black}{Initial feedback we gained from nurses and RTs informed both the interaction design and modeling work for this concept. Moving forward, we aim to iterate on the concept to convey both patient trajectory and priority –places where ICU clinicians think AI can help \cite{eini2022tell}– to help clinicians consider and perform evidence-based protocols.}

\section{Discussion}

\textcolor{black}{Our work have explored facilitating early stage AI ideation and problem formulation} – an opportune moment for involving domain stakeholders in identifying the right thing, or a good enough thing, to build \cite{passi2019problem}.
We built on the prior observation that
effective innovation teams brainstorm \textit{many} AI concepts
% designers and HCI experts can expand and facilitate participation in AI development 
by using AI capabilities and examples, before selecting a concept to further develop \cite{yang2018investigating, yildirim2022experienced, yildirim2023investigating}.
We share a case study detailing how our multidisciplinary team effectively engaged in brainstorming AI concepts for the ICU. Below, we reflect on how this approach can generalize to high-stakes, critical domains to reduce the risk of developing unwanted technology. We detail what challenges remain for moving from ideation to prototyping, and discuss open research questions and the limitations of this work.

\subsection{Towards Participatory AI in High-Stakes, Critical Domains}
% \subsection{An Emergent Design Process for AI}

\textcolor{black}{Researchers have called for participatory approaches to AI for engaging a broad set of stakeholders in early phase brainstorming to explore AI's potential value and risks in high-stakes, critical domains \cite{kawakami2022care, cai2021onboarding, delgado2022uncommon, halfaker2020ores}. However, 
% There is a growing body of literature investigating how domain experts can be meaningfully involved in high-stakes, critical domains, including child welfare \cite{kawakami2022care}, healthcare \cite{cai2021onboarding}, law \cite{delgado2022uncommon}, and content moderation \cite{halfaker2020ores}. While there is an emerging consensus that stakeholders \textit{should} participate in AI development,
it remains unclear how, when and to what extent this would be possible \cite{bratteteig2018does, birhane2022power, robertson2020if, delgado2021stakeholder}.
We took a step towards this direction in the context of healthcare.}
% Our goal was to broadly explore AI’s problem-solution space in the context of ICU \textit{before} we selected what to build. 
This is a relatively challenging design space to navigate, as we did not bind our ideation to specific AI mechanisms (e.g., clinical NLP) or interaction forms (e.g., AI-assisted diagnosis). We approached this challenge by holding design workshops, hoping that by bringing data science, HCI, and domain experts together, we could elicit what is clinically relevant and feasible. However, simply asking clinicians what would be most valuable did not prove effective: concepts were largely unbuildable or unwanted. We suspect that following a user-centered process has unintentionally led our team to focus on problems that do not need AI – points of great uncertainty or edge cases where AI is not likely to work. Additionally, traditional rules of brainstorming, such as letting go of technical limitations, seemed to exacerbate the problem of generating unbuildable concepts.

In search of a more effective process, we took a step towards matchmaking \cite{bly1999design}. Starting with AI capabilities and examples, and then asking clinicians if they recognize situations where capabilities would be useful \textcolor{black}{and where moderate performance could create value}, led to more effective ideation. It resulted in a broader coverage of the problem-solution space, leading to technically achievable and clinically relevant concepts. Capability abstractions and examples scaffolded clinicians’ understanding of what AI can do, and gave our team a shared language to discuss what would be possible. In addition to discovering value, engaging domain experts in concept generation and assessment helped us surface potential risks. We were able to identify which data features we should \textit{not} use, data that could not be trusted.

This provides a glimpse into what effective ideation and problem formulation might look like, and how it might help situate AI in high-stakes, critical work contexts. Future research should investigate whether this approach might generalize within and beyond healthcare. Does reviewing AI capabilities and examples \textcolor{black}{with moderate performance} help domain experts systematically yield high-impact, low-risk concepts? How does the selection of examples and capabilities impact the quality of generated concepts? Comparing the two brainstorming approaches – workshop 1 and workshop 2 – poses additional challenges: it is difficult to assess whether there is an interaction or order effect, since starting with AI capabilities will immediately sensitize the team to what AI can do. We encourage HCI and design researchers to share first-person accounts and case studies swapping and modifying these approaches to ideation to guide our community in constructing a better design process as well as new educational exercises. Recent work (e.g., \cite{murray2023grasping, murray2022metaphors}) provides great starting places for this line of inquiry.

Our work focused on clinicians as the domain stakeholders, yet there are many critical stakeholders in healthcare including patients, caregivers, hospital managers, insurance companies, and regulatory bodies. How could we blend matchmaking with participatory design where all stakeholders can meaningfully engage? What is the earliest point in the design and development process to engage domain stakeholders? While we started our project post-collection, we suspect that generating AI concepts \textit{prior} to data collection could inform the collection of high quality data in the first place. Recent literature suggested proactive and intentional data collection practices through pre-collection planning and documentation \textcolor{black}{\cite{hopkins2023designing, pushkarna2022data, zajkac2023ground, noortman2022breaking, tan2023seat}}. Future research can build on this line of work by engaging diverse stakeholders in \textit{designing data} to inform what should and should not be collected.

\vspace{-2mm}

\subsection{Moving from Ideation to Prototyping}
\textit{Sketching} –generating many different ideas in order to discover the right thing to make– and \textit{prototyping} –making the thing at increasing levels of fidelity to refine it into being– are cornerstones of HCI practice \cite{buxton2010sketching}. Envisioning and prototyping AI experiences pose many unique challenges for innovation teams, especially at the early stages of ideation, problem formulation, and project selection \cite{yildirim2023investigating}. Throughout our ideation process, we \textcolor{black}{utilized} several resources and artifacts that can serve as a launching pad for cross-disciplinary, collective ideation. To summarize, we used:

\begin{itemize}
    \item \textit{A set of AI capability abstractions and examples, detailing what AI can do and how it has previously produced value, \textcolor{black}{especially with moderate performance.}} These capabilities offered a starting point for discussing whether AI could solve a problem that particularly seemed like a good match. The capability abstractions provided a shared language and encouraged our team to bring up more examples throughout the ideation.
    \item \textit{\textcolor{black}{A combination of assessment matrices delineating task expertise-model performance and impact-effort.}} Noticing the interplay between these dimensions helped us map the design space, and guided our search and prioritization.
    \item \textit{A worksheet capturing the interaction, model reasoning, and data.} The Do-Reason-Know worksheet enabled us to effectively enrich an concept and understand its potential impact and limitations. It helped us to separate interaction form and model behavior. It also supported a deeper discussion on the data source, allowing us to flag data features that were unavailable, unreliable, or potentially biased.
\end{itemize}

Starting our project, one of our goals was to identify \textit{low hanging fruit} – situations where simple AI interventions could improve clinical work.
\textcolor{black}{Based on prior research highlighting the value `imperfect AI' can bring \cite{bossen2023batman, kocielnik2019will} as well as our own work, we focused on \textit{AI model performance} to sensitize our team to situations where moderate model performance can still bring enough value. Additionally, we repeatedly probed team members to think of simpler versions of concepts.}
% However, we did not have great clarity on how to describe the opportunity space or how to narrow it down. Should we look for simple-to-build models or simple outputs (e.g. binary classification)? Through deliberation, \textit{AI model performance emerged} as a key dimension: How to search for situations where moderate model performance can still bring enough value?
This explicit consideration opened up a design space beyond the automation of mundane tasks or augmentation of critical tasks. It surfaced things that humans would never do as it would not be worth their time for the return value \textit{(e.g. predicting patients with high uncertainty to receive a clinical protocol, predicting what medications will be needed for patients to \textcolor{black}{reduce pharmacy wait times)}}. These low-risk situations present a great entry point for introducing AI in healthcare, which can inform our understanding of how people can and should collaborate with AI before deploying AI in high-stakes situations, such as decision making.

While these resources scaffolded and improved our ideation process, challenges remain in selecting a concept for further development. How do we analyze, compare, and select concepts in a more systematic way? Can we engage a broad set of impacted stakeholders, including patients and other clinical roles, to anticipate risks, \textcolor{black}{fairness issues, and potential harm?} What are some critical dimensions that are not captured by current assessment tools? Recent research uncovered assessment matrices industry practitioners created to assess and prioritize AI-enabled product features, which captured risk, frequency of use, and accuracy \cite{yildirim2023investigating}. Similarly, our discussions surfaced risk of errors, data quality, acceptable model performance, and timing and presentation of information as key aspects to consider. Future research should investigate developing new assessment tools that move beyond typical metrics (e.g. feasibility, desirability) to capture the complexity of \textcolor{black}{AI concept proposals.} Moving from ideation to parallel prototyping –both experience prototyping and prototyping with data– our community would benefit greatly from having a robust assessment and selection process.

\subsection{Open Research Questions}
Our study revealed several open research questions. Below, we detail two challenges that merit further study.

\subsubsection{How much AI knowledge is needed for domain experts to engage in ideation?} Recent HCI research has explored the critical role domain experts play in AI development processes, especially in high-stakes domains \cite{sambasivan2022deskilling, cheng2021soliciting}. Researchers note that AI developers cannot readily elicit input from domain experts, and are often compelled to hold AI education sessions to span communication gaps \cite{kross2021orienting, nahar2022collaboration, piorkowski2021ai}. What kind of AI literacy is needed for domain experts to effectively participate in AI envisionment?
% In our case, the resources that the design/HCI team created for building an own understanding became also resources for facilitating domain experts' understanding around AI.
What kind of AI resources can help domain experts in engaging in ideation? How can we extend the set of AI capabilities and examples for use in other domains and contexts? \textcolor{black}{Developing and assessing resources for stakeholder engagement in ideation, problem selection and formulation marks a clear direction for future research.}

\subsubsection{What makes an AI example ``good''?}
% Prior research has highlighted the importance of capability abstractions in ideating with AI \cite{yang2018investigating, yildirim2022experienced}. Our case study echoes this: capability abstractions and examples effectively scaffolded our collective ideation and problem formulation. However, a critical question emerged:
Our research surfaced a key question: what makes an example `good'? How do we select \textit{a good enough subset of examples} that illustrate a breadth of AI capabilities and value propositions? We approached the capabilities and examples only as a subset \textcolor{black}{to sensitize our team to think of other examples and capabilities. We also paid attention to the level of AI performance in each concept, and made sure to include examples where moderate performance created value.} Interestingly, our team responded well to this approach and started drawing from other examples based on each member’s prior experience. This approach of having \textit{``a good enough subset''} was effective, as it would be incredibly challenging to try to represent and go through all AI capabilities.
\textcolor{black}{In this work, we utilized the AI Brainstorming Kit \cite{yildirim2023creating} to select capabilities and examples.
Future research should investigate the use of this resource and others (e.g., \cite{jansen2023mix}) to explore how selecting a subset of capabilities impact ideation, and how teams can effectively curate and review capabilities and examples.}

\subsubsection{Can early phase ideation and assessment address the high AI failure?}
User-centered design and participatory design grew out of HCI research addressing high rates of software product failures in the 1970s and 80s. Software engineers would select applications and start writing code; the idea of investigating what users want, need, and fear \textit{before} making software was non-obvious. We see parallels between early software development and current AI product development. Recent research echoes this: industry product teams report repeatedly experiencing AI project failures due to working on the wrong problem \cite{yildirim2023investigating}. We suspect that HCI experts can play a key role in AI development by helping teams find \textit{the right AI thing to build while reducing the risk of potential harm}. This is especially true in high-stakes contexts, such as healthcare and public sector \cite{cai2021onboarding, kawakami2022care, delgado2022uncommon, cheng2021soliciting}, where AI teams do not seem to ideate on their own. HCI routinely facilitates the process of technology innovation between multiple stakeholders to reduce the risk of developing products and services nobody wants. What is uniquely difficult about facilitating AI ideation? We strongly encourage researchers to explore the role of HCI in facilitating collective AI ideation and problem formulation.

\section{Limitations}
Our study had two limitations. First, we focused solely on sketching. While we are in the process of prototyping and model building for a few of the selected concepts, we do not claim that all concepts we generated are feasible, valuable, or novel in practice. Instead, we assess the perceived difficulty and perceived value of the concepts. This trade-off between sketching and prototyping was intentional, as our focus was on broadly exploring many concepts. Future research should investigate ideation followed by parallel prototyping of multiple concepts to assess the impact and technical effort required for implementation. Second, we do not know if there was an order effect on our ideation process. Future work should conduct controlled studies to compare the user-centered \textit{and} tech-centered approach we propose with traditional, user-centered brainstorming.

\section{Conclusion}
This paper presented a case study of early phase AI innovation capturing multidisciplinary concept ideation and problem formulation \textcolor{black}{in the context of healthcare.} Our work offers insights into how teams might structure their design process to effectively explore AI’s problem-solution space and engage domain experts in ideation. We documented our case with high-fidelity, detailing the challenges we encountered and our emergent solutions. Our case suggests that starting ideation with AI capabilities leads to broader exploration of the solution space, \textcolor{black}{and sensitizing teams to the level of AI performance needed surfaces lower risk concepts where moderate AI performance can still be useful.} It also suggests that detailing concepts in terms of data, inference, and form helps to rapidly identify problems and makes concepts more pliable \textcolor{black}{to interrogate easier, simpler versions.} While we conducted this work in the context of intensive care, we suspect this ideation and problem formulation process would generalize to many AI innovation projects that involve domain experts. Through this work, we hope to deepen the discussion on HCI's role in engaging multidisciplinary teams and stakeholders in AI ideation.

\begin{acks}
We thank the participants in this work for their time and valuable
input on the dashboard concepts. This material is based upon work supported by the National Science Foundation under Grant No. (2007501) and work supported by the National Institutes of Health (R35HL144804). The first author was also supported by the Center for Machine Learning and Health (CMLH) Translational Fellowships in Digital Health. Any opinions, findings, and conclusions or recommendations expressed in this material are those of the authors and do not necessarily reflect the views of the National Science Foundation or the National Institutes of Health.
\end{acks}

\bibliographystyle{ACM-Reference-Format}
\bibliography{icu-ideation}

% \appendix









\end{document}